\documentclass[english,russian,12pt]{article}
\usepackage{amssymb,amsmath,soul,amsthm}
\usepackage[cp866]{inputenc}
\usepackage[russian]{babel}
\usepackage{mathrsfs}
\begin{document}
\begin{center}\bf{ THE MAXIMUM PRINCIPLE OF THE NAVIER - STOKES EQUATION\footnote{$^)$
The work of the Committee on Intellectual Property Rights, Ministry of Justice of the Republic of Kazakhstan
registered. CERTIFICATE  \No 039, 12.01.2011г. {\rm ИС} 0006098}$^)$} \end{center}
\centerline{\small\sc  Akysh Abdigali Shoiynbaiuly }
  \begin{center}{\small\sc Institute of Mathematics of the Ministry of Education and Science\\ of the Republic of Kazakhstan, Almaty,} akysh41@\rm{g}mail.com\end{center}

{\small \textbf{Abstract.}
 In the work of Navier-Stokes (NSE) equation, derived a nonlinear parabolic equation for kinetic energy density, and identified an important property of this equation - the maximum principle. The latter shows the validity of the maximum principle and the NSE. On the basis of what, the unique solvability of the weak and the existence of strong solutions for NSE was proved wholly in time $t\in[0,T],$ $\forall{T}<\infty$.}

 {\bf 1~Introduction.} The current state of the mathematical theory of Navier-Stokes equation appears, for example, in \cite{lad}--\cite{Akm}, etc. Major unresolved problems of the theory of the Navier-Stokes equation of a homogeneous fluid are given in \cite{lad}--\cite{FF}. In particular, in the monograph of Ladyzhenskaya (\cite{lad}; p.13), we formulate some unsolved problems in mathematical theory of Navier-Stokes equation. Apparently, among them the principal is:

  {\it 1) Is there a unique solution in general, the general three-dimensional initial boundary value problem in a class of generalized solutions without any assumptions about the smallness of known functions and areas filled with fluid?}

In several papers \cite{Ash2}-\cite{Ash15} and others of the author, some basic statements were received in order to study the maximum principle for the NSE. The system of nonlinear parabolic equation for kinetic energy density, and important property of this equation - the maximum principle was derived from NSE. With the last the validity of the maximum principle for the NSE was shown, which from a mathematical point of view is the key. In this paper, these results are summarized and linked to the mathematical rigor and, based on them the unique solvability of the weak and the existence of strong solutions to the NSE wholly in time $t\in[0,T],$ $\forall{T}<\infty$ was proved. In this paper, these results are summarized and linked to the mathematical rigor.

{\bf 2~A statement of the problem.} Consider the initial-value problem for NSE \cite{lad} regarding the velocity  ${\mathbf U}=(U_1,U_2,U_3)$ and pressure  $ P $
  in the domain  $Q=(0,T]\times \Omega $:
 \begin{subequations}\label{is1}
  \begin{eqnarray}
&& \frac{\partial\mathbf{U}}{\partial t}-\mu\Delta\mathbf{U}
 +(\mathbf{U},\nabla){\mathbf U}+\nabla P=\mathbf{f}(t,\mathbf{x}),\label{is1a}\\
&&\text{div}\mathbf{U}=0,\label{is1b}\\&&\mathbf{U}(0,\mathbf{x})=\mathbf{\Phi}(\mathbf{x}),\quad \label{is1c}\\ &&
\mathbf{U}(t,\mathbf{x})\bigl{|}_{\partial\Omega}=0,\quad\mathbf{x}\in\partial\Omega, \label{is1d}
 \end{eqnarray} \end{subequations}
   where  $\mathbf{x}\in\Omega\subset R_3;$ $\Omega$ -- the convex region filled with a homogeneous liquid, and  $\partial\Omega$ --its boundary $\Omega$, $ t\in [0,T],\,T<\infty;$
   $\mathbf{f}$ и $\mathbf{\Phi}$ -- vector-functions accordingly to the
   external forces and initial data;  $0<\mu$ -- dynamic viscosity
   coefficient; $\Delta$ и $\nabla$ --operators of Laplace and Hamilton, respectively.
  Let  $\mbox{\bf\r{J}}(\Omega)$ -- the space of solenoidal vectors, and
 ${\mathbf G}(\Omega)$ consists of  $\nabla\eta,$ where $\eta$ is single-valued function in $\Omega$.
It is known  \cite{lad}, \cite{We} orthogonal resolution ${\mathbf L}_2(Q)={\mathbf G}(Q)\oplus \mbox{\bf\r{J}}(Q),$  moreover the elements of $\mbox{\bf\r{J}}(Q)$
at $\forall t$ belong  to $\mbox{\bf\r{J}}(\Omega),$ and the elements  $\mathbf{G}(Q)$-to subspace $\mathbf{G}(\Omega);$ $W_{2,0}^k(\Omega)$ is Sobolev space of vanishing functions on $\partial\Omega$;
the following will be used -- $L_2\bigl{(}0,T; W_{p,0}^k(\Omega)\bigr)\equiv L_2(0,T)\cap{W_{p,0}^k(\Omega)};$\\
 $W_{2,0}^{2,1}(Q)$ -- the Hilbert space of space of vanishing functions on $[0,T]\times\partial\Omega$   and having generalized derivatives of $\big{\{} U_t,\> U_{x_\alpha},\>U_{x_\alpha x_\beta},{\small(\alpha,\beta=\overline{1,3})}\big{\}}$ from $L_2(Q)$;\\
Assume that input data of problem (\ref{is1}) are $\mathbf{f}$, $\mathbf{\Phi}$ satisfied following requests:\\
$${\bf i})\> {\mathbf f}(t,\mathbf{x})\in\mathbf{C}(\bar{Q})\cap\mbox{\bf\r{J}}(Q)
 ; \quad {\bf ii})\> {\mathbf\Phi}(\mathbf{x})\in \mathbf{C}(\bar{\Omega})\cap{\mathbf W}_{2,0}^1(\Omega)
\cap\mbox{\bf\r{J}}(\Omega).$$
  We will use the well-known Holder inequality
 \begin{equation} \Bigl|\int\limits_\Omega UV\,\mathbf{dx}\Bigr|\leq
 \Bigl(\int\limits_\Omega|U|^p\,\mathbf{dx} \Bigr)^\frac{1}{p}
 \Bigl(\int\limits_\Omega|V|^q\,\mathbf{dx}\Bigr)^\frac{1}{q}\label{H1} \end{equation}
and Young for the paired products
 \begin{equation} UV\leq\frac{1}{\epsilon p}|U|^p
 +\frac{\epsilon}{q}|V|^q,\quad \epsilon>0,\quad \frac{1}{p}+\frac{1}{q}=1, \label{Y1}
  \end{equation}
Moreover, the formula for integration by parts
 \begin{equation}
 \int\limits_{\Omega} V\Delta U\,\mathbf{dx}=-\int\limits_{\Omega}\nabla V\,\nabla U\,\mathbf{dx}+
 \int\limits_{\partial\Omega}V\frac{\partial U}{\partial\rm\mathbf{n}}\,\mathbf{dx}\label{ip}\end{equation}
and well-known theorem on the solvability of the Neumann problem for Poisson equation, for example,  \cite{Mh}.

 {\bf Theorem~1.} {\it  In order to have a generalized solution of problem to be existed  \begin{equation} -\Delta V= \varphi,  \quad \frac{\partial V}{\partial{\rm{\mathbf n}}}\Bigm|_{\partial\Omega}=0, \label{npn}\end{equation}
 is necessary and sufficiently that
 $$\varphi\in L_2(\Omega)\wedge\int\limits_\Omega \varphi\,\mathbf{dx}= 0. $$
In this supposition there exists a unique generalized solution of $V$, which satisfies the condition
 $$V\in W_2^1(\Omega)\wedge\int\limits_\Omega V\,\mathbf{dx}=0.$$
 Any other generalized solution  $V^\prime$ of this problem can be written in the from  $V^\prime=V+c,$ where  $c - $is an arbitrary constant.}

{\bf 3~The Principle of maximum.  }
 The vector equation  (\ref{is1a})  put $\mathbf f=0$, and multiply by the velocity vector $\mathbf{U}$,and then using the formula
\begin{equation*}
 \Delta{E}=(\mathbf{\Delta{U}},\mathbf{U})+\sum\limits_{\alpha=1}^3|\nabla{U_\alpha}|^2,\label{ms12}
   \end{equation*}
 obtain a nonlinear parabolic equation for the density of kinetic energy (k. e.)  $E=\frac{1}{2}(U_1^2+U_2^2+U_3^2)$:
 \begin{equation}
\mathbb{L}E\equiv\frac{\partial E}{\partial t}-\mu\Delta E+\mu\sum\limits_{\alpha=1}^3|\nabla U_\alpha|^2
 +(\nabla E,{\mathbf U})+(\nabla P,{\mathbf U})=0,\label{ms2a}\end{equation}
where $\mid\mathbf{U}\mid$ is a module of the velocity vector,   $(\cdot\,,\,\cdot)$-scalar product of vectors.

 {\bf Theorem 2}\cite{Ash12}.{\it Suppose  $\bar{Q}=[0,T]\times\bar{\Omega}$ - a cylindrical domain with boundary $[0,T]\times{\partial\Omega}$ in the space of variables  $t,\mathbf{x}$ and function  $(\mathbf{U},E)\in C(\bar{Q})\cap C^2(Q)\wedge P\in{C^1(Q)}$satisfy the equations (\ref{is1a}), (\ref{ms2a}).
Then the function  $E(t,\mathbf{x})$ takes its maximum in the cylinder  $\bar{Q}$ on its lower base  $\{0\}\times\bar{\Omega}$ or on lateral area $[0,T]\times\partial\Omega$ , i.e.,}
 \begin{equation}E(t,\mathbf{x})\leq\max\bigl{\{}\sup\limits_{t=0\bigwedge\mathbf{x}\in\bar{\Omega}}E(t,\mathbf{x}),
 \sup\limits_{t\in[0,T]\bigwedge{\mathbf{x}\in\partial\Omega}}E(t,\mathbf{x})\bigr{\}}=C-const.\label{ms4}\end{equation}

 {\bf Definition~1.} {\it Let's say that the vector of
velocity $\mathbf{U}(t,\mathbf{x})$ at the point
$M_1(t^\prime,\mathbf{x}^\prime)$ of domain   $Q$ extreme, if each
component of the velocity vector
 $U_\alpha(t,\mathbf{x}),$ $\alpha=\overline{1,3}$ at that point $M_1$ reaches a
 local extremum (either a local maximum or local minimum).}

    For the proof of theorem~1 we need to have auxiliary conclusion\footnote{$^)$
    Lemma will be proved in the assumptions regarding the function $E,\>P$ of Theorem 2.
    }$^)${\bf.}

{\bf Lemma~1}\cite{Ash17, Ash16}. {\it Scalar product of  a vector of speed
$\mathbf{U}$ and its derivative
$\frac{\partial\mathbf{U}}{\partial{x_\beta}}$ \textbf{directed on vector}
$\mathbf{U}$ generates derivative of density k. e.
$\frac{\partial{E}}{\partial{x_\beta}}$  on $x_\beta$, i.e.
\begin{equation}\label{mms1}
\frac{\partial{E}}{\partial{x_\beta}}=
(\mathbf{U},\frac{\partial\mathbf{U}}{\partial{x_\beta}});
\Longrightarrow
\frac{\partial{E}}{\partial{x_\beta}}=|\mathbf{U}|\,|\frac{\partial\mathbf{U}}{\partial{x_\beta}}|\cos\gamma,\>
\text{moreover} \> \cos\gamma\neq 0, \>\forall\mathbf{x}\in\Omega,\>
 \beta=\overline{1,3}, \end{equation}
 where $\gamma-$ the angle between vectors $\mathbf{U}$ and $\frac{\partial\mathbf{U}}{\partial{x_\beta}}.$}

{\bf The proof}. Vector $\mathbf{U}$,  following \cite{Kog}, we
will present in a kind
\begin{equation}\label{mms2} \mathbf{U}=|U|\mathbf{e},  \end{equation}
 where an $\mathbf{e}(\mathbf{x})-$identity vector.

From here differentiating on $x_\beta$,  we will find
\begin{equation}\label{m1}
    \frac{\partial\mathbf{U}}{\partial{x_\beta}}=\frac{\partial|U|}{ \partial{x_\beta}}\mathbf{e}+|U|\frac{\partial\mathbf{e}}{\partial{x_\beta}}.
\end{equation}
Have as a result received expansion of a derivative of vector
$\mathbf{U}$ on two components from which the first is directed on
vector $\mathbf{U}$, and the second is directed on a
perpendicular to $\mathbf{U}$.

We will multiply scalar expansion (\ref{m1}) by the vector $\mathbf{U}$
and taking account of (\ref{mms2}), we will write down in a kind
\begin{equation*}
    (\mathbf{U},\frac{\partial\mathbf{U}}{\partial{x_\beta}})=(|U|\mathbf{e},\frac{\partial|U|}{ \partial{x_\beta}}\mathbf{e})+(|U|\mathbf{e},|U|\frac{\partial\mathbf{e}}{\partial{x_\beta}}),
    \>\forall\mathbf{x}\in\Omega.
\end{equation*}

 From here \begin{equation*}
    (\mathbf{U},\frac{\partial\mathbf{U}}{\partial{x_\beta}})=\frac{\partial{E}}{\partial{x_\beta}}(\mathbf{e}, \mathbf{e})+ |U|^2(\mathbf{e},\frac{\partial\mathbf{e}}{\partial{x_\beta}}), \>\forall\mathbf{x}\in\Omega;\> \beta=\overline{1,3} .
\end{equation*}
Of which the perpendicular part of
vector$\frac{\partial\mathbf{U}}{\partial{x_\beta}}$
$\mathbf{U}$ falls out (the second summand in the right part)  and
as a result it is found \begin{equation*}
\frac{\partial{E}}{\partial{x_\beta}}=(\mathbf{U},\frac{\partial\mathbf{U}}{\partial{x_\beta}}),\>
 \beta=\overline{1,3},\> \forall\mathbf{x}\in\Omega.
\end{equation*}

 Whence we deduce that when the vector changes both on length,
 and in a direction perpendicular complement to derivative
 $\frac{\partial\mathbf{U}}{\partial{x_\beta}}$, falls out
  the scalar product, and by that $\cos{\gamma}\neq{0},$ $\forall\mathbf{x}\in\Omega$,
   Q.E.D.

 {\bf Lemma~2}\cite{Ash12}. {\it Let the density k.e.
$E(t,\mathbf{x})$ has in any point $M_1(t,\mathbf{x})$ in the
domain $\bar{Q}=[0,T]\times\Omega$ is a local maximum, then the
point  $M_1$ is a stationary point of the velocity vector
$\mathbf{U}$, i.e.
\begin{equation}
  \nabla U_\alpha(M_1)=0,\quad \alpha=\overline{1,3}.\label{mssq}\end{equation}
 and  $\mathbf{U}$ in  the point $M_1$ reaches a local  extremum. Moreover at least at this point  $M_1$   one of the components of the velocity $\mathbf{U}$ reaches a positive maximum or negative minimum, and the rest - a negative maximum (positive low).}

{\bf Proof}.\footnote{$^)$ In \cite{Ash12}, Lemma~2 is proved on the basis of the requirement that the sufficient conditions on the major minor of the matrix quadratic forms, where the function $E(\mathbf{U})$ has a local maximum.
}$^)$   We expand the composite function  $E(\mathbf{U})=\frac{1}{2}\sum\limits_{\alpha=1}^3U_\alpha^2(t,\mathbf{x})$ in the vicinity of point  $ M_1(t,\mathbf{x})$ of the Taylor formula and the records, omitting time $t$. We obtain in this case that,
  $$ \Delta{E}\equiv E(\mathbf{x+dx})-E(M_1)=dE\big{|}_{M_1}+\frac{1}{2}d^2E\big{|}_{M_1}+o(|\mathbf{dx}|^2),$$
where the symbol $o(|\mathbf{dx}|^2)$ denotes an infinitesimal function of higher order than that, $|\mathbf{dx}|^2$.

By the previous formula, we substitute the expression differentials  $dE$, $d^2E$ computed, following \cite{ISS},  and neglecting the small value, we obtain
\begin{multline}\label{mssd}
\Delta{E}=\sum\limits_{\alpha=1}^3U_\alpha\sum\limits_{\beta=1}^3\frac{\partial{U_\alpha}}{\partial{x_\beta}}\Big{|}_{M_1}dx_\beta+\\
+\frac{1}{2}\bigg{[}\sum\limits_{\beta=1}^3\sum\limits_{\gamma=1}^3\sum\limits_{\alpha=1}^3U_\alpha\frac{\partial^2U_\alpha}
{\partial{x_\beta}\partial{x_\gamma}}dx_{\beta}dx_{\gamma}
+\sum\limits_{\alpha=1}^3\Big(\sum\limits_{\beta=1}^3\frac{\partial{U_\alpha}}{\partial{x_\beta}}dx_\beta\Big)^2\bigg{]\Bigg{|}_{M_1}}.
 \end{multline}
 By hypothesis, the point $M_1$ function $E$ has a local maximum.
 Then the necessary and sufficient conditions for local extremum $dE\big{|}_{M_1}=0$ and  is $ d^2E\big{|}_{M_1}<0$ ,  i. e. in the point $M_1$ at first differential of the function $E$  is zero,  while the second is negative.

Let see the additives from (\ref{mssd})
  \begin{equation}\label{mssb}
dE=\sum\limits_{\alpha=1}^3U_\alpha\sum\limits_{\beta=1}^3\frac{\partial{U_\alpha}}
 {\partial{x_\beta}}\Big{|}_{M_1}dx_\beta=0,\quad
 S(M_1)=\sum\limits_{\alpha=1}^3\Big(\sum\limits_{\beta=1}^3\frac{\partial{U_\alpha}}{\partial{x_\beta}}\Big{|}_{M_1}dx_\beta\Big)^2.
 \end{equation}

 The first differential by $dE$ (\ref{mssb}), interchanging the sums, we rewrite
 $$ \sum\limits_{\beta=1}^3\sum\limits_{\alpha=1}^3U_\alpha\frac{\partial{U_\alpha}}
 {\partial{x_\beta}}dx_\beta\Big{|}_{M_1}=0,$$

Whence, by virtue of the arbitrariness of the differentials of the independent variables $dx_\beta$, we obtain
\begin{equation}\label{m0}
\frac{\partial{E}}{\partial{x_\beta}}\biggm{|}_{M_1}=\big{(}\mathbf{U},\>\frac{\partial\mathbf{U}}{\partial{x_\beta}}\big{)}\Big{|}_{M_1}=0, \quad\beta=\overline{1,3}.
\end{equation}

Further, noting that relation (\ref{mms1}) at the point $M_1$ coincides with the conditions (\ref{m0}), we write
\begin{equation}\frac{\partial{E}}{\partial{x_\beta}}(M_1)= \bigl{|}\mathbf{U}(M_1)\bigl{|}\,\Bigl{|}\frac{\partial\mathbf{U}}{\partial{x_\beta}}(M_1)\Bigr{|}\cos\gamma=0,\>
\beta=\overline{1,3},  \label{ms2q} \end{equation}

 where $\gamma-$ is the angle between $\mathbf{U}(M_1)$ and $\frac{\partial\mathbf{U}}{\partial{x_\beta}}(M_1).$

From (\ref{ms2q}), we conclude that $\bigl{|}\mathbf{U}(M_1)\bigl{|}\neq 0$, because $E$ has a local maximum at point $M_1$ and $\cos\gamma=1$  on the basis of proved assertion.

From this it follows that
$\frac{\partial{E}}{\partial{x_\beta}}(M_1)=0$ if and only if $\Big{|}\frac{\partial\mathbf{U}}{\partial{x_\beta}}(M_1)\Big{|}=0.$ As a result, from (\ref{ms2q}) we obtain the chain
\begin{equation*}
       \sum\limits_{\alpha=1}^3\Big{(}\frac{\partial{U_\alpha}}{\partial{x_\beta}}(M_1)\Big{)}^2=0;\>\beta=\overline{1,3}; \Rightarrow\nabla{U_\alpha}(M_1)=0, \> \alpha=\overline{1,3}.   \end{equation*}
 The first part of the lemma is proved.

 {\bf Corollary 1}\cite{Ash17}. \emph{When the function $E(\mathbf{U})$ at the point $M_1$
 has a local maximum, whereas for (\ref{m0}), it is necessary and sufficient
 that the equality}
   \begin{equation}\label{md}
    \frac{\partial\mathbf{U}}{\partial{x_\beta}}\biggm{|}_{M_1}=0,\quad\forall\beta=\overline{1,3}.
 \end{equation}
 
 {\sf The necessity.}  Let $\frac{\partial\mathbf{U}}{\partial{x_\beta}}\big{|}_{M_1}\neq0$
 and satisfy (\ref{m0}), whereas
$\mathbf{U}$ and $\frac{\partial\mathbf{U}}{\partial{x_\beta}}$
are perpendicular at the point $M_1$,
since $\mathbf{U}\neq 0$  by hypothesis Lemma~2.
Which is impossible by Lemma 1, the vector
$\mathbf{U}$   and $\frac{\partial\mathbf{U}}{\partial{x_\beta}}$  from (\ref{m0})
are not perpendicular at $\forall\mathbf{x}\in\Omega$.

 {\sf The sufficiency.} Let  $\frac{\partial\mathbf{U}}{\partial{x_\beta}}\big{|}_{M_1}=0$
 if the point  $M_1$  satisfies (\ref{m0}) and from (\ref{md}) it follows that 
$\nabla{U}_\alpha(M_1)=0,\>\alpha=\overline{1,3}.$

{\bf Remark~1.} From $\nabla{U_\alpha}(M_1)=0, \> \alpha=\overline{1,3}$ follows, that $S\Big{|}_{M_1}=0$.
It is easy to show justice of the converse that
from $$S(M_1)=0, \Longrightarrow dE(M_1)=0,$$ whence their equivalence.

Really, we will notice from (\ref{mssd}), that for performance sufficient
conditions of a local maximum  $$ d^2E\big{|}_{M_1}<0, $$
 necessary is
$$S_\alpha(M_1)=\Big(\sum\limits_{\beta=1}^3dx_\beta\frac{\partial{U_\alpha}}
{\partial{x_\beta}}\Big{|}_{M_1}\Big)^2=0,\>\forall\alpha=\overline{1,3}.$$

From here
\begin{equation*}
 \frac{\partial{U_\alpha}}{\partial{x_1}}dx_1+\frac{\partial{U_\alpha}}{\partial{x_2}}dx_2+\frac{\partial{U_\alpha}}{\partial{x_3}}dx_3\Big{|}_{M_1}=0
. \end{equation*}
From it by virtue of the arbitrariness of the independent differentials
variable $dx_\beta$ we obtain $\nabla{U_\alpha}(M_1)=0, \> \alpha=\overline{1,3}$.
Which confirms the correctness of the proof of the lemma given in
\cite{Ash12}.

{\bf Remark~2.} {\it We have proved that the stationary points of function $E(\mathbf{U})$ coincide with stationary points of the velocity components $\mathbf{U}$. However, this result is not sufficient for approval that, at least one component of the velocity at this point should reach a supremum. }

In this connection, we prove the second part of the lemma.
For this we consider a sufficient condition for local maximum $E$ in point $M_1$ with respect to the principal minors of a symmetric matrix corresponding to the second differential in  (\ref{mssd}) $$ \mathbf{E}(M_1)=
 \begin{Vmatrix}\sum\limits_{\alpha=1}^3U_\alpha\frac{\partial^2U_\alpha}
{\partial{x_1^2}}&\sum\limits_{\alpha=1}^3U_\alpha\frac{\partial^2U_\alpha} {\partial{x_1}\partial{x_2}}&
  \sum\limits_{\alpha=1}^3U_\alpha\frac{\partial^2U_\alpha}{\partial{x_1}\partial{x_3}}\\\\
\sum\limits_{\alpha=1}^3U_\alpha\frac{\partial^2U_\alpha}
{\partial{x_2}\partial{x_1}}&\sum\limits_{\alpha=1}^3U_\alpha\frac{\partial^2U_\alpha} {\partial{x_2^2}}&
  \sum\limits_{\alpha=1}^3U_\alpha\frac{\partial^2U_\alpha}{\partial{x_2}\partial{x_3}}\\\\
 \sum\limits_{\alpha=1}^3U_\alpha\frac{\partial^2U_\alpha}
{\partial{x_3}\partial{x_1}}&\sum\limits_{\alpha=1}^3U_\alpha\frac{\partial^2U_\alpha} {\partial{x_3}\partial{x_2}}&
  \sum\limits_{\alpha=1}^3U_\alpha\frac{\partial^2U_\alpha}{\partial{x_3^2}}\\
\end{Vmatrix}\quad $$
in the form: $ E_1(M_1)<0; \quad E_2(M_1)>0; \quad E_3(M_1)<0,$ and that means negative definite matrix $\mathbf{E}$, where the principal minors are denoted by  $E_\beta$ and $\beta$ indicates the order of the minor.
 Furthermore in the computation we won't $M_1$ indicate the point of entry.

We write the inequality for the first principal minor of the matrix
\begin{equation}E_1=\sum\limits_{\alpha=1}^3U_\alpha\frac{\partial^2U_\alpha}{\partial{x_1^2}}<0.\label{er1}\end{equation}
This inequality is possible if and only if at least one component of the velocity satisfies the sufficient condition for a positive maximum or negative minimum on the variable $x_1$ at the point $M_1$, and the rest - a negative maximum (positive low) and the vector $\mathbf{U}$ reaches the extreme of a variable $x_1$.

Indeed, let's in any $\alpha$ component $U_\alpha$ is satisfied the sufficient condition of positive maximum (negative minimum) in variable $x_1$, then $$
U_\alpha >0\wedge\frac{\partial^2U_\alpha}{\partial{x_1^2}} <0 \quad
\Bigl(U_\alpha <0\wedge\frac{\partial^2U_\alpha}{\partial{x_1^2}} >0 \Bigr),$$
then so that
\begin{equation}
U_\alpha \frac{\partial^2{U_\alpha}}{\partial{x_1^2}} <0.    \label{er2a} \end{equation}

 When this inequality (\ref{er2a}) be fulfilled at $\forall\alpha$, then all components $U_\alpha$ are satisfied the sufficient condition of positive maximum (negative minimum) in variable $x_1$ be fulfilled (\ref{er1}). However, this can't be in a the best  accident only one of the components can be satisfied the sufficient condition of positive maximum(negative minimum), but the rest - components are satisfied the sufficient condition of negative maximum (positive minimum) and inequality (\ref{er1}) be fulfilled, there is inequality
 \begin{equation*} \Bigl{|}U_\alpha \frac{\partial^2U_\alpha}{\partial{x_1^2}} \Bigr{|}>
\sum\limits_{\beta\neq\alpha}U_\beta \frac{\partial^2U_\beta}{\partial{x_1^2}} , \end{equation*}
i.e. module of left-hand side of inequality (\ref{er2a}) must exceed the sum of the remaining two terms in (\ref{er1}).
If this is not the case, then together with it and inequality (\ref{er1}). Then there remains the case that any two velocity components satisfy the sufficient condition for a positive maximum or negative minimum, and the last component satisfies the sufficient condition for the maximum negative (positive low) and probably the inequality (\ref{er1}). Vector $\mathbf{U}$ reaches the extreme of a variable $x_1$.

 Now consider the inequality of the principal minor of second order at:
  $$
E_2 =\begin{vmatrix} &E_1 &\sum\limits_{\alpha=1}^3U_\alpha\frac{\partial^2U_\alpha}{\partial{x_1}\partial{x_2}}\\\\
&\sum\limits_{\alpha=1}^3U_\alpha\frac{\partial^2U_\alpha}{\partial{x_1}\partial{x_2}}
&\sum\limits_{\alpha=1}^3U_\alpha\frac{\partial^2U_\alpha}{\partial{x_2^2}}\\ \end{vmatrix}>0.$$
 Whence $$ E_2 =
E_1\sum\limits_{\alpha=1}^3U_\alpha\frac{\partial^2U_\alpha}{\partial{x_2^2}}
-\Bigl(\sum\limits_{\alpha=1}^3U_\alpha\frac{\partial^2U_\alpha}{\partial{x_1}\partial{x_2}}\Bigr)^2>0.
 $$
 Taking into account (\ref{er1}), the minor $E_2$  will be positive if and only if
 $$
\sum\limits_{\alpha=1}^3U_\alpha\frac{\partial^2U_\alpha}{\partial{x_2^2}}<0. $$
Whence, arguing as in the case of (\ref{er1}), we show that the velocity vector $\mathbf{U}$
reaches an extremum  at variable $x_2$ with respect and at least one component satisfies the sufficient condition for a positive maximum (negative minimum) at variable $x_2$ with respect to (note that this is the component that reached a positive maximum (negative minimum) on $x_1$), and the rest - a negative maximum (positive minimum).

Finally, let's consider the minor of third order. Inequality for $E_3 $ to be written as
  \begin{equation*}
E_3 =\begin{vmatrix}E_1&
    \sum\limits_{\alpha=1}^3U_\alpha\frac{\partial^2U_\alpha}{\partial{x_1}\partial{x_2}}&
 \sum\limits_{\alpha=1}^3U_\alpha\frac{\partial^2U_\alpha}{\partial{x_1}\partial{x_3}}\\
 \sum\limits_{\alpha=1}^3U_\alpha\frac{\partial^2U_\alpha}{\partial{x_1}\partial{x_2}}
 &\sum\limits_{\alpha=1}^3U_\alpha\frac{\partial^2U_\alpha}{\partial{x_2^2}}&
 \sum\limits_{\alpha=1}^3U_\alpha\frac{\partial^2U_\alpha}{\partial{x_2}\partial{x_3}}\\
\sum\limits_{\alpha=1}^3U_\alpha\frac{\partial^2U_\alpha}{\partial{x_1}\partial{x_3}}
 &\sum\limits_{\alpha=1}^3U_\alpha\frac{\partial^2U_\alpha}{\partial{x_2}\partial{x_3}}&
  \sum\limits_{\alpha=1}^3U_\alpha\frac{\partial^2U_\alpha}{\partial{x_3^2}}\\
\end{vmatrix}\equiv\end{equation*} \begin{multline} \equiv\frac{1}{E_1} \begin{vmatrix}E_1&
     \sum\limits_{\alpha=1}^3U_\alpha\frac{\partial^2U_\alpha}{\partial{x_1}\partial{x_3}}\\\\
  \sum\limits_{\alpha=1}^3U_\alpha\frac{\partial^2U_\alpha}{\partial{x_1}\partial{x_3}}
  &\sum\limits_{\alpha=1}^3U_\alpha\frac{\partial^2U_\alpha}{\partial{x_3^2}}\\
\end{vmatrix}E_2-\\ -\frac{1}{E_1} \begin{vmatrix}E_1&
     \sum\limits_{\alpha=1}^3U_\alpha\frac{\partial^2U_\alpha}{\partial{x_1}\partial{x_3}}\\\\
 \sum\limits_{\alpha=1}^3U_\alpha\frac{\partial^2U_\alpha}{\partial{x_1}\partial{x_2}}
 &\sum\limits_{\alpha=1}^3U_\alpha\frac{\partial^2U_\alpha}{\partial{x_2}\partial{x_3}}\\
\end{vmatrix}^2<0.\label{er7}\end{multline}
Validity of (\ref{er7}) can be verified by direct calculation of determinants.

Taking into account signs $E_1$ and $E_2$ from (\ref{er7}) we do conclusion that $E_3$ will be negative if and only if, when
 $$ \begin{vmatrix}E_1&
     \sum\limits_{\alpha=1}^3U_\alpha\frac{\partial^2U_\alpha}{\partial{x_1}\partial{x_3}}\\
  \sum\limits_{\alpha=1}^3U_\alpha\frac{\partial^2U_\alpha}{\partial{x_1}\partial{x_3}}
  &\sum\limits_{\alpha=1}^3U_\alpha\frac{\partial^2U_\alpha}{\partial{x_3^2}}\\
\end{vmatrix}>0. $$  From here, $$E_1\Bigl(\sum\limits_{\alpha=1}^3U_\alpha\frac{\partial^2U_\alpha}{\partial{x_3^2}}\Bigr)-
\bigl(\sum\limits_{\alpha=1}^3U_\alpha\frac{\partial^2U_\alpha}{\partial{x_1}\partial{x_3}}\bigr)^2>0.$$
 This inequality is possible if and only if $$\sum\limits_{\alpha=1}^3U_\alpha\frac{\partial^2U_\alpha}{\partial{x_3^2}}<0.$$

As in previous cases, this inequality holds if and only if the velocity vector $\mathbf{U}$   reaches an extremum with respect $x_3$ to at least one component satisfies the sufficient condition for a positive maximum (negative minimum) with respect to $x_3$ (again, note that this is the component that has reached positive maximum (negative minimum) with respect $x_1$ and $x_2$), and the rest - a negative maximum (positive low). Lemma~2 is proved.

 For the pressure  function $P$ is an analogous lemma.

{\bf Lemma~3 }\cite{Ash14}. {\it If the density k.e.  $E(t,\mathbf{x})$ reaches its maximum at some point $M_1(t^\prime,\mathbf{x}^\prime)$ of domain $Q=[0,T]\times\Omega$,  then the point $M_1$ is a stationary point for the pressure function $P,$  i.e. equalities are correct  $\nabla P(M_1)=0$.}

{\bf Proof.} We  write the well-known formula of vector analysis
 \begin{equation} (\mathbf{U},\nabla)\mathbf{U}=\nabla E-[\mathbf{U},\mathbf{\omega}],\quad{\omega}=\rm{rot}\mathbf{U}. \label{ms8} \end{equation}
where $[\cdot,\cdot]-$ vector product.

Under the hypotheses of Theorem~2 the vector-function $[\mathbf{U},\omega]$
is continuous in a bounded domain $\Omega$ for all $t\in[0,T]$, then so that
  $[\mathbf{U},\omega]\in \mathbf{L}_2(\Omega)$, $\forall{t}\in[0,T]$. Then,
  following \cite{lad}, vector-function $[\mathbf{U},\omega]$ in the form of an  orthogonal sum
  \begin{equation}[\mathbf{U},\mathbf{\omega}]=\nabla{R}+\mathbf{V}^{(\mathbf{J})},
 \>\>\text{where}\>\> \nabla{R}\in\mathbf{G}(Q),\> \mathbf{V}^{(\mathbf{J})}\in\mbox{\bf\r{J}}(Q).
 \label{ms9}\end{equation}

  Where, at the same time, we calculate the bounded condition for $R$
  \begin{equation}\frac{\partial{R}}{\partial\rm\mathbf{n}}\Bigm|_{\partial\Omega}=0.
 \label{ms10} \end{equation}
 because  $\mathbf{V}^{({\mathbf J})}{\rm{\mathbf n}}\Bigm|_{\partial\Omega}=0$ and
$[\mathbf{U},\mathbf{\omega}]{\rm{\mathbf n}}\Bigm|_{\partial\Omega}=0$
correspondingly by virtue of $\mathbf{V}^{(\mathbf{J})}\in \mbox{\bf\r{J}}(\Omega)$ and (\ref{is1d}),
where ${\rm{\mathbf n}}$ is the identity vector of external normal in the point $x$ of boundary $\partial\Omega.$
The validity of relation (\ref{ms9}) is followed from solvability of corresponding problem of Neumann
for $R$ with right part ${\rm div}[\mathbf{U},\omega]$ and bounded condition (\ref{ms10}) on  based Theorem~1.

Applying the operator $\it{div}$ on the vector-function $(\mathbf{U},\nabla)\mathbf{U}$,  we find
\begin{equation} \text{div}\bigl((\mathbf{U},\nabla)\mathbf{U}\bigr)=\sum\limits_{\alpha,k=1}^{3}\frac{\partial
U_\alpha}{\partial x_k}\frac{\partial{U_k}}{\partial x_\alpha}. \label{mi1} \end{equation}
And formula (15) taking into account the expansion (16) rewrite
 \begin{equation}
(\mathbf{U},\nabla)\mathbf{U}=\nabla E-\nabla{R}-\mathbf{V}^{(\mathbf{J})}. \label{bmi} \end{equation}
 Apply to both sides (19) operation $\it{div}$ , and using (18), we obtain
 \begin{equation}
 \sum\limits_{\alpha,\beta=1}^{3}\frac{\partial U_\alpha}{\partial x_\beta}\frac{\partial{U_\beta}}{\partial x_\alpha} = \Delta{E}-\Delta{R}.\label{ms10a} \end{equation}
Hence, for the point $M_1$ of maximum  function $E$ we find
  \begin{equation*}
 \Delta{E}(M_1)-\Delta{R}(M_1)=0,
 \end{equation*}
 since the left side of (\ref{ms10a}) vanishes at a point $M_1$ on the basis of (\ref{mssq}).

Whence $\Delta{R}(M_1)\leq 0$, since under the hypotheses of Lemma $\Delta{E}(M_1)\leq 0$, i.e. for a function $R$ in point $M_1\in Q$ at the necessary condition of a local maximum. Thus, the point $M_1$ is a stationary point and for the function $R(t,\mathbf{x})$. From which it follows that $\nabla{R}(M_1)=0. $

Next, equation (\ref{is1a}), we multiply the gradient of an arbitrary single-valued function
 $\eta(t,\mathbf{x})\in L_\infty\bigl(0,T; C^\infty(\Omega)\bigr)$.
And then, using the orthogonality of subspaces $\mathbf{G}(\Omega)$, $
 \mbox{\bf\r{J}}(\Omega)$, we integrate over the domain $\Omega$, and as a result we
\begin{equation*} \int\limits_\Omega\Bigl(\nabla{P}
+(\mathbf{U},\mathbf{\nabla})\mathbf{U}\Bigr)\nabla\eta\,\mathbf{dx}=0, \quad \forall t\in[0,T].\label{ip3}
\end{equation*}

 Hence, by replacing the integrand $(\mathbf{U},\mathbf{\nabla})\mathbf{U}$  corresponding value of formula (\ref{bmi}) and, given the orthogonality of subspaces $\mathbf{G}(\Omega)$, $ \mbox{\bf\r{J}}(\Omega)$, we find
 \begin{equation*}\int\limits_{\Omega}\nabla(P+E-R)\nabla\eta\mathbf{dx}=0,\quad \forall{t}\in[0,T]. \end{equation*}
 Where, due to the arbitrariness $\nabla{\eta}$, we have
   \begin{equation*}
 \nabla{P}(t,\mathbf{x})+\nabla{E}(t,\mathbf{x})-\nabla{R}(t,\mathbf{x})=0,\quad \forall{(t,\mathbf{x})}\in
 Q.\label{ms15}\end{equation*}
  This identity can be written the point $M_1\in Q$ of maximum function $E$
 \begin{equation*}
\nabla{P(M_1)}+\nabla{E(M_1)}-\nabla{R(M_1)}=0.
 \end{equation*}
 Whence $ \nabla{P(M_1)}=0$, since $\nabla{E(M_1)}=0$ and $\nabla{R(M_1)}=0$ at the point $M_1\in Q$ of maximum function $E$, or extreme velocity $\mathbf{U}$. Lemma~3 is proved

 {\bf Proof of Theorem~2.} For this we use the well-known method (\cite{VS}; p.511). Assume the contrary, i.e. the function $E(t,\mathbf{x})$ reaches its maximum value at some point $M_0(t^0,\mathbf{x}^0)$ within the  domain $Q=(0,T]\times\Omega$ .
 \begin{equation}
E(M_0)>\max\bigl{\{}\sup\limits_{t=0\bigwedge\mathbf{x}\in\bar{\Omega}}E(t,\mathbf{x}),
 \sup\limits_{t\in[0,T]\bigwedge{\mathbf{x}\in\partial\Omega}}E(t,\mathbf{x})\bigr{\}}=C\geq 0.
\label{ms16}\end{equation}
 Denote $m=E(M_0)-C>0$ and introduce $H(t,\mathbf{x})=E(t,\mathbf{x})+\frac{m}{2}(1-\frac{t}{T})$. The function
$H(t,\mathbf{x})$ also takes its maximum value at some point $M_1\in Q$, and $H(M_1)\geq H(M_0)\geq m$.
Now, using results of lemmas~1,~2 we'll copy all necessary conditions of maximum of function $H$  in point $M_1$
 \begin{equation} \frac{\partial H}{\partial t}\geq 0;\quad\Delta H\leq 0;\Longrightarrow\bigl{\{}\nabla{U_\alpha}=0,\>
\alpha=\overline{1,3};\quad\nabla{H}=0;\quad \nabla P=0 \bigl{\}}. \label{ms16a} \end{equation}
From the equation (\ref{ms2a}), using the conditions (\ref{ms16a}), for the point $M_1$, we can find a chain of inequalities
\begin{equation*}\mathbb{L}H(M_1)\equiv\frac{\partial H}{\partial
t}-\mu\Delta{H}+\mu\sum\limits_{\alpha=1}^3(\nabla{U_\alpha})^{2}+(\nabla{H},\mathbf{U})+
(\nabla{P},\mathbf{U})+\frac{m}{2T}\geq\frac{m}{2T}>0. \label{ms17}\end{equation*}
This means that inequality (\ref{ms16}) is false. Consequently, we have (\ref{ms4}). Theorem~2 is proved.
Theorem 2 and Lemmas 1 and 2 allow the following maximum principle for equation (\ref{is1a}):

{\bf Corollary~2.}{ \it
Let $\bar{Q}=([0,T]\times\bar{\Omega})-$ the cylindrical domain in the space of variables $\>t,\mathbf{x}$ with boundaries of $[0,T]\times\partial\Omega$ and the function $ \mathbf{U}\in C(\bar{Q})\cap C^2(Q)\wedge{P}\in{C^1(Q)}$ satisfy the equation (\ref{is1a}).Then the vector-function $\mathbf{U}$ attains a local extremum in the cylinder  $\bar{Q}$ on the lower ground $\{0\}\times\bar{\Omega}$ or on its side surface $[0,T]\times\partial\Omega$ and at least one of the function ${U_\alpha}$ reaches a positive maximum or negative minimum, i.e.}
   \begin{subequations}\label{ms19}
 \begin{eqnarray}
&& U_\alpha(t,\mathbf{x})\leq\max\bigl{\{}\sup\limits_{t=0\bigwedge\mathbf{x}\in\bar{\Omega}}U_\alpha(t,\mathbf{x}),
 \sup\limits_{t\in[0,T]\bigwedge{\mathbf{x}\in\partial\Omega}}U_\alpha(t,\mathbf{x})\bigr{\}},\quad
 (t,\mathbf{x})\in\bar{Q}
 \label{ms19a}\\
&&\Bigl(U_\alpha(t,\mathbf{x})\geq
\min\bigl{\{}\inf\limits_{t=0\bigwedge\mathbf{x}\in\bar{\Omega}}U_\alpha(t,\mathbf{x}),
 \inf\limits_{t\in[0,T]\bigwedge{\mathbf{x}\in\partial\Omega}}U_\alpha(t,\mathbf{x})\bigr{\}},\quad
 (t,\mathbf{x})\in\bar{Q}\Bigr),
\label{ms19b} \end{eqnarray} \end{subequations} where  $\alpha=\overline{1,3}. $

 {\bf The proof } follows from Theorem 2 and Lemma 2, since the lemma, starting from the implementation of the necessary and sufficient condition for local maximum $E$, and therefore also true. Whence (\ref{ms19}).

Hence, following (\cite{VS}; p. 513), it is easy to obtain proof of the following statements:

{\bf Corollary~3.}{\it If the vector-function  $ \mathbf{f}$, $\mathbf{\Phi} $ satisfy the condition {\bf i)} and {\bf ii)}, then  for the solution  $\mathbf{U}(t,\mathbf{x})$  of problem (\ref{is1}) estimate is correct:}

 \begin{equation}\|\mathbf{U}\|_{\mathbf{C}(\bar{Q})}\leq
\|\mathbf{\Phi}\|_{\mathbf{C}(\bar{\Omega})}+T\|\mathbf{f}\|_{\mathbf{C}(\bar{Q})}\equiv A_1,\>
 \forall T<\infty, \> \|\mathbf{U}\|_{\mathbf{C}(\bar{Q})}=\max\limits_{1\leq\alpha\leq
 3}\sup\limits_{\bar{Q}}|U_\alpha(t,\mathbf{x})|.\label{ns13}\end{equation}

 {\bf 4~Weak generalized solution. }
 We multiply equation (\ref{is1a}) by an arbitrary vector-function
$\mathbf{Z}(t,\mathbf{x})\in\mathbf{C}(\bar{Q})\cap\mathbf{W}_2^1(Q)\cap\mbox{\bf\r{J}}(Q)$, equaled to zero at
$\bigl(t=T\bigr)\wedge\bigl(\mathbf{x}\in\partial\Omega\bigr)$. We shall integrate product on domain
$Q=[0,T]\times\Omega$ and with the help of an integration by parts  (\ref{ip}) from  the first two summands we shall transfer from $\mathbf{U}$ to $\mathbf{Z}.$ As a result, shall receive
     \begin{multline}
   \int\limits_{Q}\Bigl(-\mathbf{U}\,\frac{\partial\mathbf{Z}}{\partial t}+
   \mu\sum\limits_{k=1}^3\nabla U_k\nabla Z_k
       +(\mathbf{U},\nabla)\mathbf{U}
  \mathbf{Z}\Bigr)\,\mathbf{dx}\,dt=\\ \int\limits_\Omega\mathbf{\Phi}\mathbf{Z}(0,\mathbf{x})\mathbf{dx}
+\int\limits_{Q}\mathbf{f}\mathbf{Z}\mathbf{dx}\,dt. \label{ns7}\end{multline}

Again, equation (\ref{is1a}), we multiply the gradient of an arbitrary single-valued function $\eta\in L_2(0,T;W_2^1(\Omega))$. And then integrate over the domain $Q$, using the orthogonality of subspaces, in the end we find the identity
 \begin{equation} \int\limits_{Q}\nabla P\,\nabla\eta\,\mathbf{dx}\,dt=-\int\limits_{Q}
(\mathbf{U},\nabla)\mathbf{U}\,\nabla\eta\,\mathbf{dx}\,dt. \label{np12}\end{equation}

 {\bf Definition~2}.\footnote{$^)$ Here, thanks to the principe of maximum, weak solution be regarded in more the restricted class of function, than
in \cite{lad}.}$^)$
 {\it  We shall call as the weak generalized solution a full initial boundary
  value problem of the Navier-Stokes equations (\ref{is1})   vector-function ${\mathbf U}$ and function $P$  from space
 \begin{multline}\mathbf{U}\in\mathbf{C}(\bar{Q})\cap\mathbf{L}_\infty\bigl(0,T;\mathbf{L}_2(\Omega)\bigr)\cap
\mathbf{L}_2\bigl(0,T;\mathbf{W}_{2,0}^1(\Omega)\bigr)\cap\mbox{\bf\r{J}}(Q);\\ P\in
L_2\bigl(0,T;W_2^1(\Omega)\bigr)\wedge\bigl(\int\limits_\Omega P\mathbf{dx}=0,\,\forall t\in [0,T]\bigr)
\label{nsp}\end{multline} and satisfying the identities (\ref{ns7}), (\ref{np12}) for any
\begin{equation*}
 \mathbf{Z}(t,\mathbf{x})\in\mathbf{C}(\bar{Q})\cap\mathbf{W}_2^1(Q)\cap\mbox{\bf\r{J}}(Q)
 \wedge\bigl(\mathbf{Z}\Bigm{|}_{(t=T)\wedge(\mathbf{x}\in\partial\Omega)}=0\bigr);
 \quad \eta(t,\mathbf{x})\in L_2\bigl(0,T;W_2^1(\Omega)\bigr).
\end{equation*}}
For the validity of this definition, all integrals, incoming to  (\ref{ns7}) and (\ref{np12}),
 must be finite for any $\mathbf{Z}$,  $\eta $ from the indicated classes.

 {\bf Lemma~4.}
  {\it If the input data of problem (\ref{is1})  satisfy the requirements
 {\bf i}), {\bf ii}), then for weak generalized solution of problem  (\ref{is1}) ) the following estimates are valid:
 \begin{equation}
 \bigl{\|}\mathbf{U}\bigr{\|}_{\mathbf{L}_\infty(0,T;\mathbf{L}_2(\Omega))}^2\leq
 2\bigl{\|}\mathbf{\Phi}\bigr{\|}_{\mathbf{L}_2(\Omega)}^2
 +4T^2\bigl{\|}\mathbf{f}\bigr{\|}_{\mathbf{L}_\infty(0,T;\mathbf{L}_2(\Omega))}^2\equiv A,
\label{ns12}\end{equation} \begin{multline} \sum\limits_{k=1}^3\int\limits_0^t\bigl{\|}\nabla
U_k(\tau)\bigr{\|}^2_{\mathbf{L}_2(\Omega)}d\tau
\leq\frac{1}{\mu}\bigl{\|}\mathbf{\Phi}\bigr{\|}_{\mathbf{L}_2(\Omega)}^2+\\
\frac{2T^2}{\mu}\bigl{\|}\mathbf{f}\bigr{\|}^2_{\mathbf{L}_\infty(0,T;\mathbf{L}_2(\Omega))}\equiv A_2, \quad \forall
t\in(0,T],\label{ns14}\end{multline} \begin{equation} \Bigl{\|}\nabla P\Bigr{\|}^2_{\mathbf{L}_2(Q)}\leq
\Bigl{\|}\bigl(\mathbf{U},\nabla\bigr)\mathbf{U}\Bigr{\|}^2_{\mathbf{L}_2(Q)}\leq 9A_1^2A_2\equiv A_3.\label{ni1}
\end{equation}}

 {\bf Proof}.\footnote{$^)$The analogous estimates   (\ref{ns12}), (\ref{ns14}) are all-known, for example, from \cite{lad1}.}$^)$
Multiply scalar equation (\ref{is1a}) by the vector-function $2\mathbf{U},$
product integrate over the domain $\Omega$ and with help of
integration by parts, we transform a second term. As a  result
we obtain
    \begin{multline}
   \frac{d}{dt}\int\limits_\Omega\sum\limits_{k=1}^3|U_k|^2\,\mathbf{dx}+
   2\mu\int\limits_\Omega\sum\limits_{k=1}^3\bigl|\nabla U_k\bigr|^2\,\mathbf{dx}
   +\\
    2\int\limits_\Omega\bigl(({\mathbf U},\nabla){\mathbf U}+\nabla P\bigr){\mathbf U}\,\mathbf{dx}
 = 2\int\limits_\Omega\sum\limits_{k=1}^3 f_k U_k\,\mathbf{dx}, \quad t\in(0,T].
\label{ns11}\end{multline}

 As a consequence of the orthogonality of subspaces ${\mathbf G}(\Omega)$ and
$\mbox{\bf\r{J}}(\Omega)$ find the relation
\begin{equation*} 2\int\limits_\Omega\bigl((\mathbf{U},\nabla)\mathbf{U}+ \nabla
P\bigr)\mathbf{U}\,\mathbf{dx}=\sum\limits_{k=1}^3\int\limits_\Omega\mathbf{U}\nabla{U_k^2}\mathbf{dx}+
2\int\limits_\Omega\nabla P\mathbf{U}\mathbf{dx}=0, \quad \forall t\in[0,T].\end{equation*}

Taking into account the last identity,(\ref{ns11})integrate over $t$  a range from $0$ to $t$. Right-hand side can be estimated by Young's inequality with $\epsilon=\frac{1}{2T}$. As a result, we obtain for the energy norm $\mathbf U$
  \begin{multline}
 \bigl{\|}{\mathbf U}(t)\bigr{\|}_{{\mathbf L}_2(\Omega)}^2
+2\mu\int\limits_{0}^t\sum\limits_{k=1}^3\bigl\|\nabla U_k(\tau)\bigr\|^2_{{\mathbf L}_2(\Omega)}d\tau
 \leq \bigl{\|}{\mathbf\Phi}\bigr{\|}_{{\mathbf L}_2(\Omega)}^2
 +\\
0.5\bigl{\|}{\mathbf U}\bigr{\|}_{{\mathbf L}_\infty((0,T];{\mathbf L}_2(\Omega))}^2+ 2T^2\bigl{\|}{\mathbf
f}\bigr{\|}_{{\mathbf L}_\infty((0,T];{\mathbf L}_2(\Omega))}^2, \quad\forall t\in(0,T]. \label{ns11a}\end{multline}
Hence we have the estimate (\ref{ns12}) for the squared norm of the function $\mathbf U$. Again using (\ref{ns12}), from (\ref{ns11a}) we find that inequality (\ref{ns14}).

For proof (\ref{ni1}) in the identity (\ref{np12}) we put $\nabla\eta=\nabla P$, and  then
  estimate the right-hand part at Young inequality (\ref{Y1}) at $p=2\wedge\epsilon=1$
  and as a result we will have the inequality
    \begin{equation*}
\bigl{\|}\nabla P\bigr{\|}^2_{\mathbf{L}_2(Q)}\leq\bigl{\|} (\mathbf{U},\nabla)\mathbf{U}\bigr{\|}^2_{\mathbf{L}_2(Q)}.
\end{equation*}
 The right-hand part of the last we estimate successively on Cauchy-Bunyakovsky inequality
 for vector product and  Holder inequality (\ref{H1}) at $p=\infty\wedge q=1$.
 In a result we have the chain
 \begin{multline*}
\Bigl{\|}(\mathbf{U},\nabla)\mathbf{U}\Bigr{\|}^2_{\mathbf{L}_2(Q)}\leq
3\int\limits_Q\bigl|\mathbf{U}\bigr|^2\sum\limits_{k=1}^{3} \bigl|\nabla U_k\bigr|^2\,\mathbf{dx}\,dt\leq\\
9\max\limits_k\bigl{\|}U_k\bigr{\|}^2_{L_\infty(Q)}
\sum\limits_{k=1}^{3}\int\limits_{0}^{T}\bigl\|\nabla{U_k(t)}\bigr\|^2_{\mathbf{L}_2(\Omega)}dt,
 \label{ns16}\end{multline*}
 from that, on the basis of estimates (\ref{ns13}),(\ref{ns14}), it follows that (\ref{ni1}). Lemma~4 is proved.

From the principle of maximum and obtained
a priori estimates, the uniqueness weak solutions of problems (\ref{is1})  are followed:

{\bf Theorem~3}\cite{Ash11,Ashk}. {\it If input data $\mathbf{f}$ and  $\mathbf{\Phi}$
satisfying requirements {\bf i}) and {\bf ii}),then
 each problem has the unique weak generalized solution $\mathbf{U}$ and  $P$ satisfying to identities  (\ref{ns7}), (\ref{np12})  at any $\mathbf{Z}$ and $\eta $ from the definition~2.}

{\bf Proof.} Let couple of function $\{\mathbf{U}, P\}$ and $\{\mathbf{U}^{*}, P^*\}$ - two solutions of problems (\ref{is1}). Put $\mathbf{V=U-U^*},\,$ $R= P-P^*$, them have:
 \begin{subequations}\label{ns19}
  \begin{eqnarray}
 && \frac{\partial\mathbf{V}}{\partial t}-\mu\Delta\mathbf{V}
+(\mathbf{V},\nabla)\mathbf{U}+(\mathbf{U^*},\nabla)\mathbf{V}+\nabla R= 0,\label{ns19a}\\\nonumber\\
&&\mathbf{V}(0,\mathbf{x})=0,\quad\mathbf{V}(t,\mathbf{x})\bigl{|}_{\partial\Omega}=0,\quad\mathbf{x} \in\partial\Omega,
\label{ns19b}
 \end{eqnarray} \end{subequations}

 From equation (\ref{ns19a}) we pass to identity
 \begin{equation}
 \int\limits_{Q_t}\bigl(\frac{\partial\mathbf{V}}{\partial t}\mathbf{V}-\mu\Delta\mathbf{V}
   \mathbf{V}
+(\mathbf{V},\nabla)\mathbf{U}\mathbf{V}+ (\mathbf{U^*},\nabla)\mathbf{V}\mathbf{V}+\nabla
R\mathbf{V}\bigl)\mathbf{dx}d\tau= 0,\quad \forall t\in(0,T].
 \label{ns20}\end{equation} $\mathbf{U}\in\mbox{\bf\r{J}}(Q)$,
So $\mathbf{U}\in\mbox{\bf\r{J}}(Q)$, that $\mathbf{V}\in\mbox{\bf\r{J}}(Q)$.
 When by virtue orthogonality of subspace   $\mbox{\bf\r{J}}(Q) $ and $\mathbf{G}(Q)$, we obtain the relation
  \begin{displaymath}
      \int\limits_{Q_t}(\mathbf{U^*},\nabla)\mathbf{V}\mathbf{V}\mathbf{dx}=0,\quad
      \int\limits_{Q_t}\nabla R\mathbf{V}\mathbf{dx}=0,\quad \forall t\in(0,T],
      \end{displaymath}
    all the other terms transform with help integration at parts(\ref{ip}), then
 from (\ref{ns20}) find
\begin{equation}\frac{1}{2}\bigl{\|}\mathbf{V}(t)\bigr{\|}^2_{\mathbf{L}_2(\Omega)}+
\mu\sum\limits_{k=1}^{3}\int\limits_0^t\|\nabla V_k(\tau)\|_{\mathbf{L}_2(\Omega)}^2d\tau=
-\int\limits_{Q_t}\sum\limits_{k,\beta=1}^{3}V_\beta\frac{\partial V_k} {\partial x_\beta}U_k\mathbf{dx}d\tau.
\label{ns21}\end{equation}
The integral in right-hand part we estimate successively on  Holder's inequality (\ref{H1})
at $p=\infty\wedge q=1$ and Young (\ref{Y1})  at $p=2$, as a result put the chain of inequality
 \begin{multline*}
\Bigl|\int\limits_{Q_t}\sum\limits_{k,\beta=1}^{3}V_\beta\frac{\partial V_k} {\partial
x_\beta}U_k\mathbf{dx}d\tau\Bigr|\leq \max\limits_k\|U_k\|_{L_\infty(Q)}
\sum\limits_{k,\beta=1}^{3}\int\limits_{Q_t}\Bigl|\frac{\partial V_k} {\partial
x_\beta}\Bigr|\bigl|V_\beta\bigr|\mathbf{dx}d\tau \leq\\ A_1\epsilon/2\sum\limits_{k,\beta=1}^{3}
\int\limits_0^t\Bigl\|\frac{\partial V_k} {\partial x_\beta}\Bigr\|^2_{L_2(\Omega)}d\tau+A_4
\int\limits_0^t\sum\limits_{\beta=1}^{3}\bigl\|V_\beta\bigr\|^2_{L_2(\Omega)}d\tau\leq\\
A_1\epsilon/2\sum\limits_{k=1}^{3}\int\limits_0^t\bigl\|\nabla V_k(\tau)\bigr\|^2_{\mathbf{L}_2(\Omega)}d\tau
+A_4\int\limits_0^t\bigl{\|}\mathbf{V}(\tau)\bigr{\|}^2_{\mathbf{L}_2(\Omega)}d\tau,\quad A_4=3A_1/(2\epsilon).
\end{multline*}
Taking into account estimates (\ref{ns13}), (\ref{ns14}) and
use the last  inequality at   $\epsilon=2\mu/A_1$  from (\ref{ns21}), we will find
\begin{displaymath}\bigl{\|}\mathbf{V}(t)\bigr{\|}^2_{\mathbf{L}_2(\Omega)} \leq
A_4\int\limits_0^t\bigl{\|}\mathbf{V}(\tau)\bigr{\|}^2_{\mathbf{L}_2(\Omega)}d\tau, \quad A_4=3A_1^2/(4\mu),\quad\forall
t\in(0,T].\end{displaymath}

From here, we have
\begin{equation}
\frac{d}{dt}\Bigl(\exp(-A_4t)\bigl{\|}\mathbf{V}(t)\bigr{\|}^2_{\mathbf{L}_2(\Omega)}\Bigr)\leq 0, \quad\forall
t\in(0,T].\label{ns22}\end{equation}
From inequality (\ref{ns22}) conclude, that
 ${\mathbf V} \equiv 0,$ $\forall t\in(0,T],$ i.e. that solution ${\mathbf U}$ and ${\mathbf U}^*$
 coincided.
Now with the help of the functional equation (\ref{np12}), considering only that the uniqueness $\mathbf{U}$, we obtain the integral relation for
 $\nabla{R}$ $$ \int\limits_Q \nabla{R}\nabla{\eta}\mathbf{dx}dt=0.$$
  Hence,
 thanks $\forall\nabla\eta$,we obtain $\nabla{R}\equiv{0}$, i.e. the pressure $P$  gradient from the definition~2 is the only way in terms of vector-function $\mathbf{U}$. Theorem 3 is proved.

 {\bf 5~Strong solution.}

 {\bf Definition~3.} {\it If in the domain  $Q$ the weak generalized solution of initial-boundary
value problem for the equations of Navier-Stokes has the every possible generalized derivatives of the
same order, as the equations this solution is called as strong.}

{\bf Theorem~4}\cite{Ash10}. {\it If input data of problem (\ref{is1}) satisfy requirements  {\bf i}), {\bf ii}) and $\partial\Omega\in{C^2}$ then each problems (\ref{is1}) has unique strong generalized solution $\mathbf{U}$ and $P$ from spaces
 \begin{equation*} \mathbf{U}\in\mathbf{W}_{2,0}^{2,1}
(Q)\cap\mbox{\bf\r{J}}_\infty(Q);\quad P\in L_2\bigl(0,T;W_2^2(\Omega)\bigr) \wedge\bigl(\int\limits_\Omega
P\mathbf{dx}=0,\forall t\in[0,T]\bigr), \end{equation*}
  satisfying to equations (\ref{is1a})  almost everywhere in $Q$, and for them estimations take place:
\begin{equation} \bigl{\|}\mathbf{U}_t\bigr{\|}^2_{\mathbf{L}_2(Q)}\leq
\mu\sum\limits_{k=1}^{3}\bigl{\|}\nabla\Phi_k\bigr{\|}^2_{\mathbf{L}_2(\Omega)}+
 5A_3+2T\bigl{\|}\mathbf{f}\bigr{\|}^2_{\mathbf{L}_\infty(0,T;\mathbf{L}_2(\Omega))}\equiv A_5,\,\,
 \label{ni2}\end{equation}
\begin{equation} \bigl{\|}\Delta\mathbf{U}\bigr{\|}^2_{\mathbf{L}_2(Q)}\leq A_5/\mu^2\equiv A_6,\quad \label
{ni3}\end{equation}
 \begin{equation} \bigl\|\nabla U_k\bigr\|^2_{\mathbf{L}_\infty(0,T;\mathbf{L}_2(\Omega))}\leq
A_5/\mu\equiv A_7, \quad k=\overline{1,3}, \label{ni3a} \end{equation}
 \begin{equation} \bigl\|\nabla
P\bigr\|^2_{\mathbf{L}_\infty(0,T;\mathbf{L}_2(\Omega))}\leq 3A_1^2A_7\equiv A_{10}, \label{ni3b} \end{equation}
\begin{equation} \|\mathbf{U}\|_{\mathbf{L}_2(0,T;\mathbf{W}_2^2(\Omega))}\leq
A_8\|\Delta\mathbf{U}\|_{\mathbf{L}_2(Q)},\quad A_8-const, \label{ni4} \end{equation}
\begin{equation}\|P\|_{L_2(0,T;W_2^2(\Omega))}\leq A_p \|\Delta P\|_{L_2(Q)}\leq
A_c\|\mathbf{U}\|_{\mathbf{L}_2(0,T;\mathbf{W}_2^2(\Omega))}, \quad A_c,A_p-const. \label{ni5} \end{equation}}

  {\bf Proof. }For the proof of inequality  (\ref{ni2}) from equations (\ref{is1a}) we will
find identity
   \begin{equation}
\int\limits_{Q_t}\bigl( \mathbf{U}_t- \mu\Delta\mathbf{U}\bigr)^2\mathbf{dx}\,d\tau = \int\limits_{Q_t}\bigl(
\mathbf{f}-(\mathbf{U},\nabla)\mathbf{U}-\nabla P\bigr)^2\mathbf{dx}\,d\tau. \label{ng5}\end{equation}
We will build a square integrand expression. After this pair product in the left-handed part of the transform by integration by parts (\ref{ip}), and the right-handed part of any efforts to Young's inequality (\ref{Y1}) at $\epsilon=1 \wedge  p=2$. And then from (\ref{ng5}) becomes the inequality
 \begin{multline*}
\int\limits_{Q_t}\mathbf{U}_t^2\mathbf{dx}\,d\tau+
\mu^2\int\limits_{Q_t}\bigl(\Delta\mathbf{U}\bigr)^2\mathbf{dx}\,d\tau
+\mu\sum\limits_{k=1}^{3}\int\limits_\Omega\bigl|\nabla U_k\bigr|^2\mathbf{dx}\leq\\
\mu\sum\limits_{k=1}^{3}\int\limits_\Omega\bigl|\nabla\Phi_k\bigr|^2\mathbf{dx}
 +5\int\limits_{Q_t}\Bigl((\mathbf{U},\nabla)\mathbf{U}\Bigr)^2\mathbf{dx}\,d\tau
+2\int\limits_{Q_t}\mathbf{f}^2\mathbf{dx}\,d\tau. \end{multline*}

From the last inequality with regard for (\ref{ni1}) we receive estimates
 (\ref{ni2})--(\ref{ni3a}) the strong generalized solutions of problem (\ref{is1}). And  (\ref{ni3a})
 is a better estimate than  (\ref{ns14}).

Equation (\ref{is1a}) we multiply the gradient of an arbitrary function  $\eta\in L_\infty(0,T;W_2^1(\Omega))$ and
integrate on domain $\Omega$
\begin{equation*} \int\limits_{\Omega}\nabla
P\,\nabla\eta\,\mathbf{dx}=-\int\limits_{\Omega} (\mathbf{U},\nabla)\mathbf{U}\,\nabla\eta\,\mathbf{dx}.
\end{equation*}
 Location, setting $\nabla\eta=\nabla{P}$, go to the inequality
 \begin{equation*} \int\limits_\Omega\mid\nabla{P}\mid^2\mathbf{dx}\leq
\int\limits_\Omega\mid\mathbf{U}\mid^2\sum\limits_{k=1}^3\mid\nabla{U_k}\mid^2\mathbf{dx}. \end{equation*}
 Right-hand part,
evaluated by the Holder inequality (\ref{H1}), at $p=1\wedge q=\infty$, we have \begin{equation*}
\int\limits_\Omega\mid\nabla{P}\mid^2\mathbf{dx}\leq 3\bigl{\|}\mathbf{U}(t)\bigr{\|}^2_{\mathbf{L}_\infty(\Omega)}
\int\limits_\Omega\sum\limits_{k=1}^3\mid\nabla{U_k}\mid^2\mathbf{dx},\>\forall t\in{[0,T]}. \end{equation*}

Hence, using (\ref{ni3a}) we arrive at (\ref{ni3b}).

Since the boundary of domain $\partial\Omega\in C^2$ we find the estimate (\ref{ni4}),  using inequality from $\bigl($\cite{lad}; p.26$\bigr)$, valid for any function $U(x)\in{ W}_2^2(\Omega)\cap{ W}_{2,0}^2(\Omega)$:
 \begin{equation*}
\|\mathbf{U}\|_{\mathbf{W}_2^2(\Omega)}\leq A_8\|\Delta\mathbf{U}\|_{\mathbf{ L}_2(\Omega)},\quad \quad \forall
t\in[0,T],\quad A_8-const.\label{lg1} \end{equation*}

Find an estimate for $\Delta{P}$ from relation
 \begin{equation*} -\Delta{P}=\sum\limits_{\alpha,\beta=1}^{3}\frac{\partial U_\alpha}{\partial
 x_\beta}\frac{\partial{U_\beta}}{\partial x_\alpha}.
\end{equation*}
found from the vector equation (\ref{is1a}) with the operation of div based (\ref{is1b}) and (\ref{mi1}).

We square both parts the last equation and we will integrate on domain  $\Omega$. Then, estimating in a right-hand part, we obtain a inequality
 \begin{equation}
 \int\limits_\Omega\bigl(\Delta P\bigr)^2\mathbf{dx}\leq
 9\sum\limits_{\alpha,\beta=1}^{3}\int\limits_\Omega\Bigl|\frac{\partial
 U_\alpha}{\partial x_\beta}\Bigr|^4\mathbf{dx},
\quad \forall t\in[0,T].\label{lg2} \end{equation}

 Owing to embedding theorems of Sobolev we have
${W}^2_2(\Omega)\subset{W}^1_{6-\epsilon}(\Omega),\,\forall\epsilon>0$.
From here when $\epsilon=2$ following inequality
\begin{displaymath}
 \|U_\alpha\|_{W_4^1(\Omega)}\leq A_9 \|U_\alpha\|_{W_2^2(\Omega)},\quad \forall t\in[0,T],
\end{displaymath}
where $A_9$--is vague constant.

On the  basis of the last inequality and (\ref{ni3}), (\ref{ni4}) from (\ref{lg2})
we will find an estimate (\ref{ni5}).

 The vector-function $\mathbf{U}$ and function of pressure $P$ subjects to estimates (\ref{ni2})-(\ref{ni5}) satisfies the equations (\ref{is1a}) almost everywhere in $Q$.
Theorem~4 is proved.

 {\bf Remark~3.} Theorem~3 about uniqueness of weak generalized solutions problems (\ref{is1}) are valid for their strong and classical solutions.

\newpage
 \renewcommand{\refname}{\begin{center}\normalsize\bf References  \end{center}}

\end{document}